\documentclass[submission,copyright,creativecommons]{eptcs}
\providecommand{\event}{FAVO 2011} % Name of the event you are submitting to
\usepackage{breakurl}             % Not needed if you use pdflatex only.
\usepackage{graphicx}
\title{Towards a Formal Model of Privacy-Sensitive Dynamic Coalitions}
\author{Sebastian Bab
\institute{Technische Universit\"at Berlin, Germany}
\email{sebastian.bab@tu-berlin.de}
\and
Nadim Sarrouh
\institute{Technische Universit\"at Berlin, Germany}
\institute{SOAMED Graduate College}
\email{n.sarrouh@tu-berlin.de}
}
\def\titlerunning{Formal Modeling of Privacy-Sensitive Dynamic Coalitions}
\def\authorrunning{Sebastian Bab \& Nadim Sarrouh}
\begin{document}
\maketitle

\begin{abstract}
The concept of dynamic coalitions (also virtual organizations) describes the temporary interconnection of autonomous agents, who share information or resources in order to achieve a common goal. Through modern technologies these coalitions may form across company, organization and system borders. Therefor questions of access control and security are of vital significance for the architectures supporting these coalitions.

In this paper, we present our first steps to reach a formal framework for modeling and verifying the design of privacy-sensitive dynamic coalition infrastructures and their processes. In order to do so we extend existing dynamic coalition modeling approaches with an access-control-concept, which manages access to information through policies. Furthermore we regard the processes underlying these coalitions and present first works in formalizing these processes. As a result of the present paper we illustrate the usefulness of the \emph{Abstract State Machine (ASM)} method for this task. We demonstrate a formal treatment of privacy-sensitive dynamic coalitions by two example ASMs which model certain access control situations. A logical consideration of these ASMs can lead to a better understanding and a verification of the ASMs according to the aspired specification.
\end{abstract}

\section{Introduction}
Modern technologies like service-oriented architectures, online social networks, web 2.0, etc.\ allow for companies and organizations to form dynamic coalitions across their own system borders. A dynamic coalition first off is a group of through network technologies (temporary) interconnected autonomous agents, who share their resources or information in order to reach a common goal. Although each coalition exhibits unique features there exist certain common features like dynamically changing membership, data-transfer mechanisms and authorization structures. Companies cooperating in order to exploit a temporary market chance or the cooperation of emergency services, military and civilian institutions during a crisis are popular examples for dynamic coalitions.

In most scenarios different institutions have to share and transfer critical information, which raises significance for questions of information flow, security, privacy and trust. In security and privacy critical fields like in the health sector the solving of the question may determine the success of complete software engineering projects. Nevertheless software architects today still lack formal frameworks and methods to evaluate policy decisions at design time and thereby examine their consequences for the to-be-deployed architecture.

In \cite{Bryans0601} Bryans et al. present formal modeling technologies in order to approach a formal framework for dynamic coalitions. With the specification language VDM\footnote{Vienna Development Method: \url{http://www.vdmportal.org/twiki/bin/view}} they create a basic formal model of dynamic coalitions, consisting of various dimensions, where each dimension represents a certain perspective which could be considered by a software architect. According to object oriented paradigms the models consist of data types as well as operations and invariants over these types. In \cite{Bryans06} the authors demonstrate the practicability of their modeling approach.

In other works independent of dynamic coalition researches the same authors present concepts for modeling Access-Control-Policies in VDM. Therefor they translate core components of the OASIS\footnote{OASIS is a consortium for advancing open standards for information society. See: \url{http://www.oasis-open.org/}.}-standard XACML (\emph{eXtensible Access Control Markup Language} into a VDM-Notation. In an outlook they propose the combination of this approach with the modeling of dynamic coalitions \cite{Bryans07}.

\begin{figure}[h]
	\centering
		\includegraphics[width=0.8\textwidth]{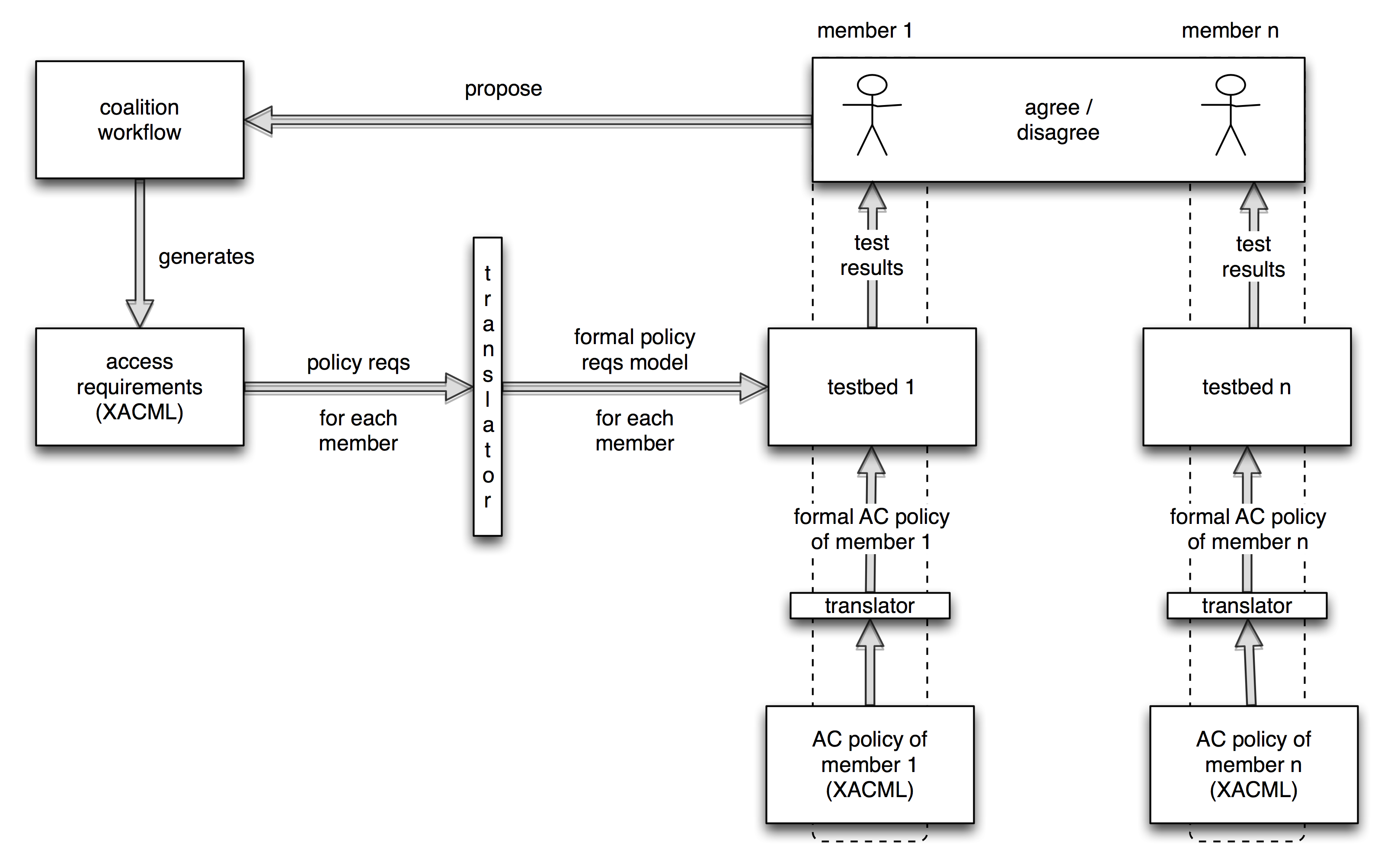}
	\caption{Evolution of access control policies in dynamic coalitions taken from \cite{Bryans07} }
	\label{fig:evoAC}
\end{figure}

Modeling and testing of proposed software architectures with respect to their access control policies is of great importance, especially in dynamic coalitions: Different access control policies of certain agents may result in processes that do not run smoothly or prevent the processes from running at all. Bryans et al. therefor present a concept of \emph{evolution of access control policies in dynamic coalitions} (see figure \ref{fig:evoAC}). In this concept a process with certain access control requirements is generated through coalition consensus. With the supposed formal framework agents may investigate these access control requirements and even simulate the impact on their own architecture, thereby being able to detect and avert errors or contradictions. 

Since one of the authors is affiliated with the SOAMED Graduate College\footnote{For more Information on SOAMED see: \url{http://www.informatik.hu-berlin.de/forschung/gebiete/soamed-en/willkommen-en}.}, a graduate college dedicated to "Service-oriented Architectures for the Integration of Software-based Processes exemplified by Health Care Systems and Medical Technology" medical scenarios are of great interest to the authors. In health care systems dynamic factors of coalitions are of great importance. Possibilities of modern networking technologies call for new organizational concepts apart from traditional static concepts like clinical pathways. For example, patient care itself could be seen as a dynamic coalition:

\begin{quote}
During the treatment in one or several hospitals the patient has to pass through various station and organizational units. These departments are often not only spatially separated, but also technology wise, acting as separate, autonomous instances, which create there own reports and diagnoses which are transferred to the other instances (up to this day still mostly on paper).
\end{quote}

In this paper we present our approach of integrating the above-described modeling concepts for access control and dynamic coalitions. Speaking in terms of Bryans et al. we thereby examine the dimension access control. The models of Bryans et al. solely address structural aspects of dynamic coalitions. Exceeding these structural features the consideration of coalition processes is of high significance in dynamic coalitions. In the core of this paper we discuss how \emph{Abstract State Machines (ASMs)} may be utilized to enrich the models of Bryans et al. with formal process aspects. The modified VDM-model is considered as the underlying state and processes may be represented as state transitions in the ASM. We provide two examples of access control situations in dynamic coalitions in the formal method of ASMs which shall illustrate the usefulness of a formal treatment and consideration of access control situations. By the verification of certain liveness and safety properties we can furthermore guarantee or check specific properties of the ASMs in the first place.

\section{Related Work}
In recent years several formalizations of single aspects of dynamic coalitions have been presented. Haidar et al. propose a formal model for PKI-based authentication in dynamic coalitions on the basis of a process calculus and the formal description language $Z$ \cite{Haidar09}. Bocchi et al. present formal description approaches for breeding environments in virtual organizations, which may be notably relevant in grid-computing.

As mentioned above, Bryans et al. contributed various models for dynamic coalitions by means of the specification language VDM (see \cite{Bryans06,Bryans07,Bryans08,Bryans0601}) . However  those approaches only cover structural aspects of dynamic coalition. Process properties and workflows have to be simulated using external (for example java-based) tools to modify a state that is based on a VDM-model structure. Our work extends this approach in so far as in that we pursue a formal modeling framework in which both structure and process properties can be formalized in a single formalism (e.g. \emph{Abstract State Machines}) and thereby create the means to formally analyze dynamic coalitions with the underlying processes.

Other than the taxonomies of Bryans et al. only few holistic modeling approaches for dynamic coalition exist. The work most alike to ours can be found in \cite{McGinnis09}. Here McGinnis et al.\ describe a formal framework for virtual organizations in service-grid-environments. The framework catches formal descriptions of agents, services, roles and work flows. The used modeling techniques are not based on standards that we know of. Tool-support in creation and evaluation of dynamic coalitions design, as proposed by us, is therefor only hardly imaginable. Koshutanski et al. model HDC (highly dynamic coalitions), a subclass of dynamic coalitions, defined by them as coalitions with extremely short life time. They also do not make use of any modeling standard, which might create problems in critical application fields like crisis management or the health sector. Furthermore their model assumes a central management of the coalition in form of a coalition platform. Both limitations are too restrictive for the application scenarios we are envisioning. 

Other approaches present informal models from the view of economic sciences. \cite{Klueber98,Lethbridge01}.

\section{Basics}
In the following we want to describe the basic concepts which are essential for the understanding of the model as to be presented in this paper. In the first subsection we give a short introduction into the modeling language VDM. In the second subsection we provide the model for dynamic coaliation as described in \cite{Bryans06}. In the third part we give a brief introduction to the OASIS XACML standard, for which there also exists a VDM specification.

\subsection{VDM and VDM++}
The model to be introduced in the present paper is specified in the object-oriented extension of the Vienna Development Method (VDM), called VDM++. This language specifies data, states and functionality and is well suited for the modeling of the structures and functions of dynamic coalitions.\cite{FitzgeraldLarsen2009,Fitzgerald2005}.

A VDM-Model consists of definitions of classes in which instance variables, types and functionality are specified. Here instance variables are the local variables of an object. Types can be defined as being free of structure (called \emph{token}) or as complex. Functionality can be specified through functions over the instance variables or through helping functions, which leave the local variables untouched. The use of operations and functions can be limited through the statement of pre- and post-conditions. VDM is widely used method in the modeling of computer-related systems \cite{Larsen96,Fitzgerald06} and adequately supported by tools like for example the VDMTools which were used for the present paper.

\subsection{Dynamic coalitions in VDM}
The VDM model for dynamic coalitions as introduced by Bryans et al.\ consists basically of agents, which can enter or leave certain groups of agents, which are named coalitions. Agents carry a set of information which can be shared with other agents or coalitions. Based on these conceptions \cite{Bryans0601} introduces certain different dimensions of dynamic coalitions for the modeling of seven further aspects like for example exchanging of information, memberships of coalitions and structures of authorization. The practical usefulness of these models was proven by certain industrial (\cite{Bryans06}) and military (\cite{Bryans08}) case studies.

In \cite{BabSarrouh11} we extended this basis model of dynamic coalitions by methods for a handling of access-control. In this respect our approach can be seen as another new dimension of dynamic coalitions: the dimension of access-control.

\subsection{XACML in VDM}
\begin{figure}[h]
	\centering
		\includegraphics[width=0.6\textwidth]{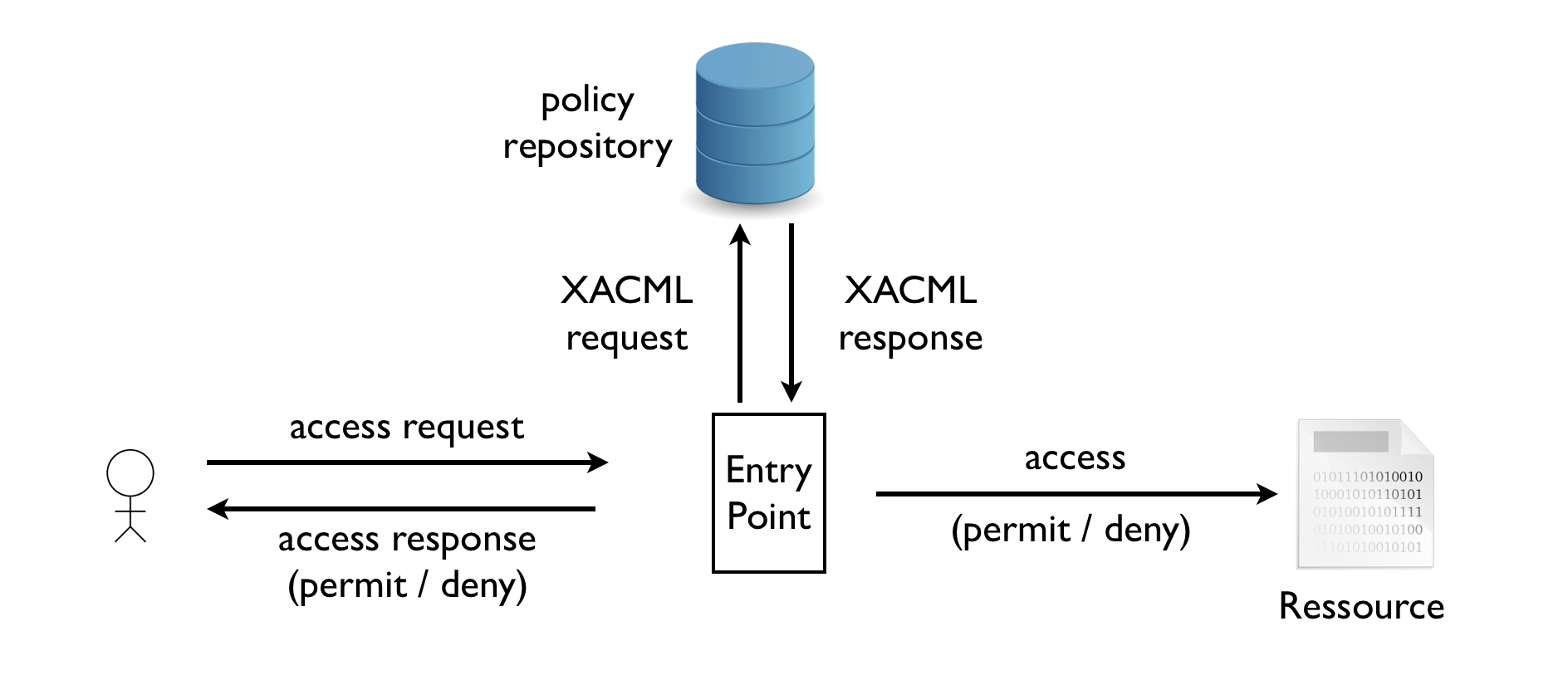}
	\caption{Simplified presentation of XACML-architecture.}
	\label{fig:xacml}
\end{figure}

XACML is an OASIS standard based on XML for the description and definition of access-control-policies in distributed systems. A simplified visualization of the functioning of XACML policies can be found in figure \ref{fig:xacml}. Requests for access to certain resources are picked up by the Policy Enforcement Point (PEP) and send to the Policy Decision Point (PDP) as a standardized XACML request. As holder of the policies it is up to the PDP who check whether an access is to be permitted or to be denied and sends this information to the PEP which is responsible for the granting or denying of access to the resource.

\subsection{Abstract State Machines}
The concept of abstract state machines (ASMs) (see for example \cite{Boe03,Gure00}) allows for a formal representation of computable and non-computable algorithms. Here the concept of ASMs is rather free as it allows for a direct use of specification languages (like for example VDM) as underlying languages of the ASMs. ASMs offer a mathematical framework which allows for a reliable formal modeling of processes. Any ASM is based on the concept of states in which the ASM can occur. Here any state is modeled as a specification of the underlying specification language. State transitions are modeled over rules which change the definitions of the functions included in the specification. This change takes place via an \emph{update} operation, which overwrites the result of the function for a certain set of arguments. Thus a state is determined by the definition of its included functions.

ASMs allow for a logical handling in the sense that one defines a logic for AMSs in which formulas describe certain states and conditions of the ASM. As a result of the present work we will show in the next sections how two examples for a formal representation of access-control questions in dynamic coalitions can be modeled and studied using the ASM method and its logical considerations.

\section{Dimension Access Control: Our VDM state}

In the following we present our basic state which is used for the ASM modeling approach in the next chapter. The models are loosely based on the work of Bryans et al. \cite{Bryans0601a} although it has been modified and extended on several locations. At first we introduce the basic types of the model, separated in types for the modeling of dynamic coalitions and access control respectively. In the last subsection we explain some example operations on the basis of which simple access controlled information transfer within a coalition becomes possible. \footnote{The full model including a test scenario can be found under \url{http://www.user.tu-berlin.de/nsarrouh/ACDimension.vpp}}

\subsection{Dynamic coalition types}

According to the approaches of Bryans et al.\ our basis state signature is based on \emph{agents} which carry information and may join forces to form dynamic coalitions and thereby sharing information at their demand. Both, \emph{agents} and \emph{coalitions} have unique IDs (\emph{Aid}- and \emph{Cid-token}). The model abstracts from the actual content and structure of the information and considers information as unstructured data (\emph{token-type}). Furthermore we add PDPs, both at agent and at coalition level, which host the access control policies for later evaluation.
\begin{verbatim}
public Agent       ::  info   : set of Information
                       aac    : PDP;
                
public Coalition   ::  agents : set of Aid
                       info   : set of Information
                       cac    : PDP;
               			
public Information ::  item   : token;

instance variables

coals : map Cid to Coalition := {|->};
agents: map Aid to Agent := {|->};
inv forall c in set dom coals & 
        (coals(c).agents subset dom agents)
\end{verbatim}

The invariant (\emph{inv}) states that only known agents may join coalitions. The PDP of a coalition and the PDPs of the agents are related as follows: As soon as an agent shares information with the coalition, the related access control policies are attached to it and stored in the coalitions PDP. This way access control requirements of the agents may be enforced on coalition level.

\subsection{Access control component types}

We integrate the approach from \cite{Bryans0601a}: According to the OASIS-standard XACML policies consist of \emph{rules} which in turn consist of an optional \emph{target}) and an \emph{effect}. Targets consist of \emph{subjects}, in our case agents, which want to access certain \emph{resources}, in our case information. \emph{Actions} define the type of access in question, which we limit to \emph{read} and \emph{write} for sake of simplicity. When an access (\emph{request}) matches a rule target, the \emph{effect} is being returned which may be \emph{permit}, \emph{deny} or \emph{not applicable} in case the target of the rule does not match the request. If a rule doesn't contain a target it will be evaluated for each request.

\begin{verbatim}
public Rule    ::  target    : [Target]
                   effect    : Effect;
\end{verbatim}
\begin{verbatim}
public Target  ::  subjects  : set of Aid
                   resources : set of Information
                   actions   : set of Action;
\end{verbatim}
\begin{verbatim}
public Request ::  target    : Target;
\end{verbatim}
\begin{verbatim}
public Action = (<WRITE> | <READ>);
public Effect = (<PERMIT> | <DENY> | <NOTAPPLICABLE>);
\end{verbatim}

Policies are sets of rules including a combination algorithm which combines different effects in case that more than one rule matches a request. Here we only consider the basic XACML combining algorithms \emph{deny overrides}) and \emph{permit overrides}. Policies have obligatory targets, so that only matching requests are evaluated. The effect of a policy is returned to the \emph{Policy Decision Point (PDP)}, which combines the different effects to only one effect, using the above-mentioned combining algorithm. Each request for access on information is evaluated through the PDP which ensures policy compliance.

\begin{verbatim}
public PDP    ::  policies      : set of Policy
                  policyCombAlg : CombAlg;
\end{verbatim}
\begin{verbatim}
public Policy ::  target        : Target
                  rules         : set of Rule
                  ruleCombAlg   : CombAlg; 
\end{verbatim}
\begin{verbatim}
public CombAlg = (<DENYOVERRIDES> | <PERMITAOVERRIDES>);
\end{verbatim}

\subsection{Operations}

The model operations \emph{CreateEmptyCoalition}, \emph{CreateNewAgent} and \emph{Join} create empty coalitions, new agents carrying information, and join agents to coalitions. Agents in coalitions may share information (method \emph{shareInfo}) in which case the matching policies are added to the coalitions PDP.

\begin{figure}[h]
	\centering
		\includegraphics[width=\textwidth]{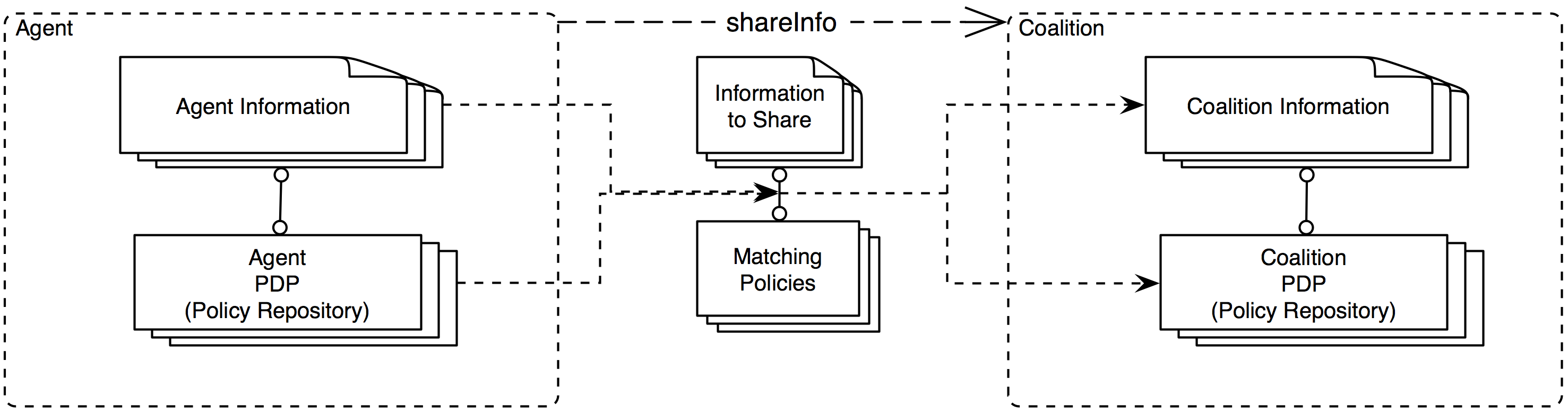}
	\caption{Visualization of \emph{shareInfo} operation.}
	\label{fig:shareinfo}
\end{figure}

\begin{verbatim}
public shareInfo : Aid * Cid * set of Information ==> ()

ShareInfo(a,c,i_set)==
( 
for all i in set i_set do
     coals:= coals ++ {c |-> mu(coals(c),
	      info |-> coals(c).info union i_set, 
          cac |-> mk_PDP(coals(c).cac.policies union
              GetMatchingPolicies(agents(a).aac, i),
                  <DENYOVERRIDES>))};
)
pre a in set dom agents and c in set dom coals and 
     i_set subset agents(a).info
;
\end{verbatim}

If an agent wants to access information, the PDP will at first evaluate the request and then permit or deny the access.

\begin{verbatim}
public RequestInfo : Aid * Cid *  Action * set of
     Information ==> Effect
RequestInfo(a,c,act,i_set)==
(
     evaluatePDP(mk_Request(mk_Target({a},i_set,{act})),
	      coals(c).cac)
)
pre a in set dom agents and c in set dom coals and 
     i_set subset coals(c).info
;
\end{verbatim}

The function \emph{evaluatePDP} checks the combining algorithm for the policies and calls the according functions. The single policies and rules are evaluated through \emph{evaluatePolicy} and \emph{evaluateRule} which in turn use the according function depending of the combining algorithm to evaluate the effects. For the sake of space we refrain from an explicit representation of all the functions and operations in this paper and refer to the complete model descriptions in the internet.

\section{Abstract State Machines}
Abstract state machines (ASMs) offer a formalism for the modeling of arbitrary algorithms (see for example \cite{Gure00,Boe03}). ASMs come with different forms of presentation of algorithms, including technical and visualized presentation forms. Furthermore there exists a tool support which allows for the actual programming with ASMs (see for example the CoreASM project). ASMs offer methods for a representation of algorithms which are not necessarily limited to computer based algorithms, but instead of that allow for the representation of other algorithmic procedures of the real world, too. The essential advantage of ASMs for the purposes of this work lies in the freedom of using nearly any modeling language and formalism as underlying formalism of a certain ASM. Thus it is possible to extend formalisms like VDM by a meta-level consideration which allows for example to reason about a VDM model from a logical perspective.

\subsection{ASM Example}
We now try to illustrate the applicability of the ASM method through a simple example for access control in a dynamic coalition taken from \cite{Bryans07}. This example is chosen due to the fact that it includes the essential questions and aspects of an access control in a dynamic coalition, while still being of an adequate simplicity. Assume that a document management tool administers production plan and hazard analysis documents of a chemical plant named \emph{compA}. Production orders for chemicals are stored in an order database and each order has to undergo identical production processes. Due to market opportunities \emph{compA} outsources the hazard analysis to \emph{compB}, thereby forming a kind of dynamic coalition. 

\begin{figure}[h] \label{fig:example1}
	\centering
		\includegraphics[width=1.0\textwidth]{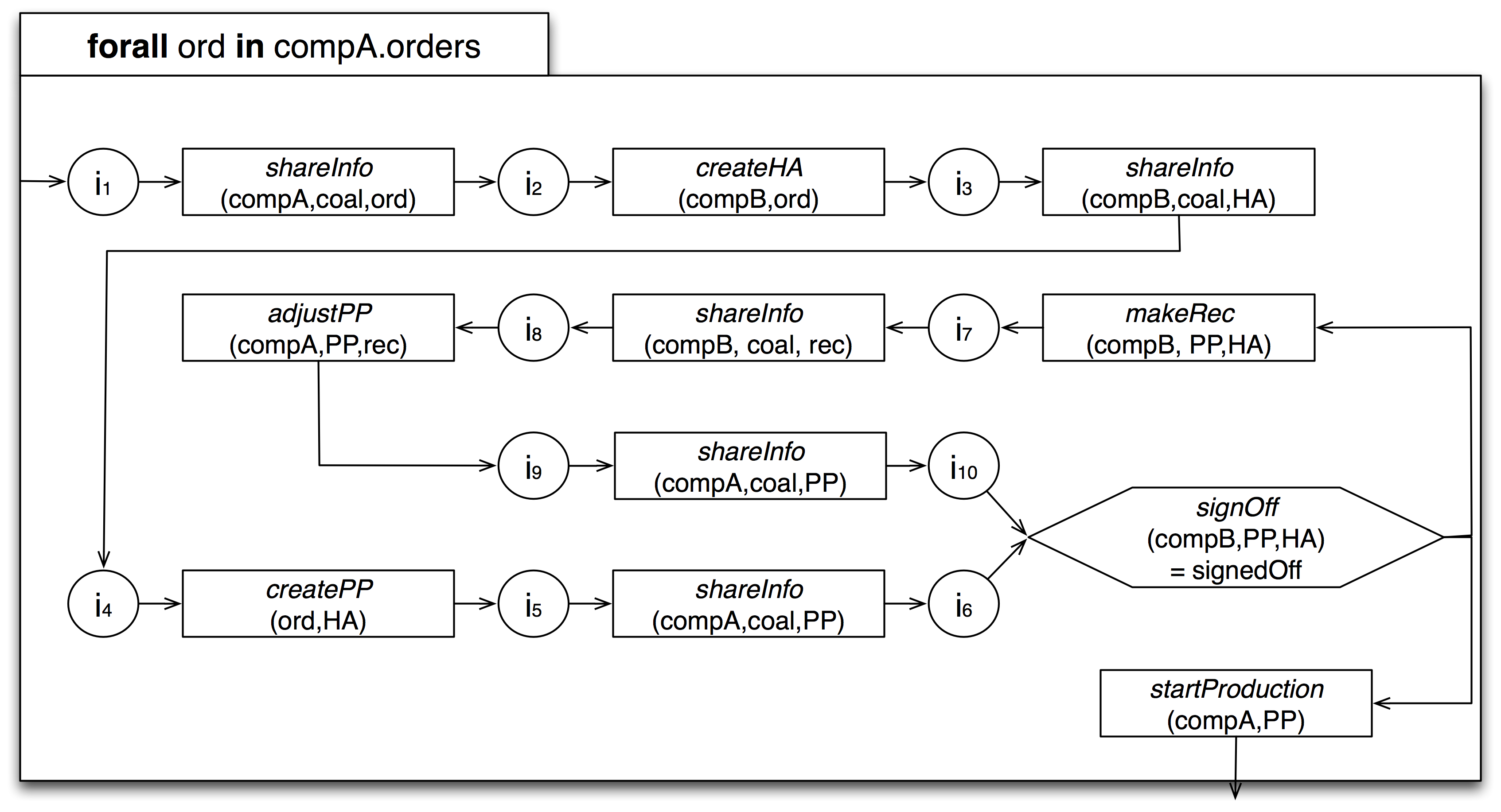}
	\caption{ASM visualization of chemical plant process with possible deadlock after i6.}
	\label{fig:proDead}
\end{figure}

Figure \ref{fig:proDead} visualizes this process in the standard ASM terminology as introduced in \cite{Boe03,Gure00}. The underlying structure is the above-mentioned VDM++-Model. According to common ASM-illustrations we refrain from defining all functions and variables explicitly in this paper and count on the expressive power of function and variable names. Round entities represent the abstract states, i.e. the whole VDM++-structure, including instances for the agents, coalitions, etc.\ whereas rectangles illustrate updates of these states. hexagons are symbolic representations of conditions. If-then-else-conditions have two possible outputs "yes" and "no" depending on the Boolean expression inside the hexagon. The process for the pre-production of a new chemical will start with the creation of a hazard analysis. According to the hazard analysis and its recommendations the production plan is developed. Suppose that legal requirements demand that \emph{compB} signs off that all security recommendations of the hazard analysis are properly implemented in the production plan.

Now suppose that company legislation and therefore the access control policy of \emph{compA} forbids any other agent but \emph{compA} to access the production plan, which is reflected by the following two rules\footnote{In order to let ruleA1 be effective the policy will have to use the PermitOverrides combining algorithm.}:

\begin{verbatim}

ruleA1:= ( Target(  {compA}  ,  {PP}  ,  {<WRITE>,<READ>}  )  ,  <PERMIT> );
ruleA2:= ( Target(  {}      ,  {PP}  ,  {<WRITE>,<READ>}  )  ,  <DENY>   );

\end{verbatim}

Because all shared information is stored in the coalition together with the according access control policies, the depicted process will run into a deadlock as soon as \emph{compB} tries to access the production plan in the first condition (we suppose that \emph{signoff()} will internally make use of the \emph{requestInfo()}-function). An ASM run easily identifies this deadlock. The resulting error will now have to be reported to the coalition members who will now have to adjust their process or their security policies in order to make the process executable. In the following process we insert another condition, checking if access to the production plan is granted. If not an error message to the participating members will result and thereby we eliminate the previous deadlock.

\begin{figure}[h] \label{fig:example2}
	\centering
		\includegraphics[width=1.0\textwidth]{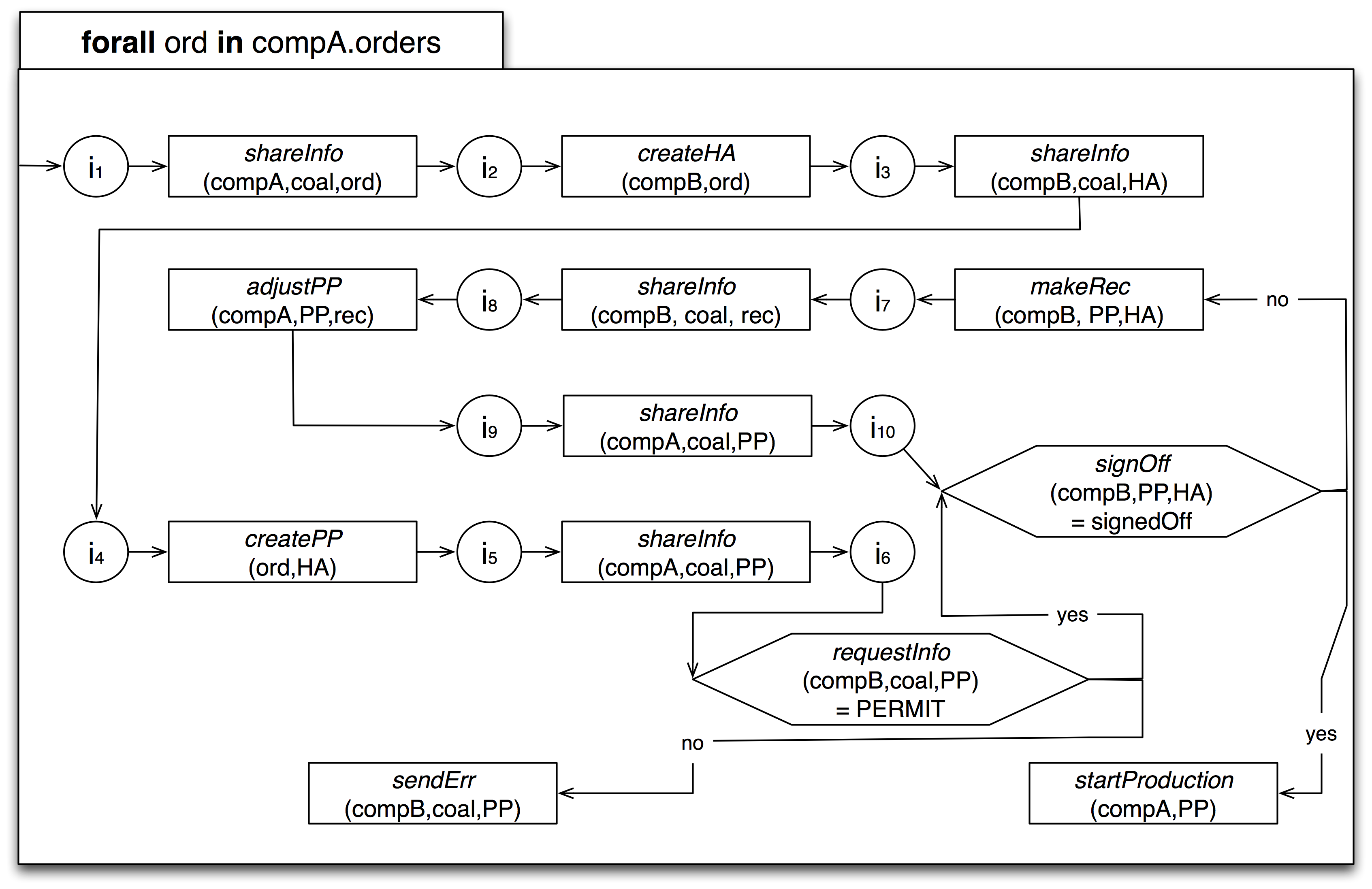}
	\caption{ASM visualization of chemical plant process with new choice to avoid deadlock after i6.}
	\label{fig:proDead2}
\end{figure}

\section{Logical considerations -- Liveness and safety}
The general method of ASMs allows for a logic based analysis of properties of the abstract state machines (compare with \cite{Boe03,Gure00}). Based on the example ASMs of the previous section we now want to argue that certain important properties in the modeling of processes like for example \emph{liveness} and \emph{safety} can be described by logical formulas over the underlying specifications of the ASMs. Here liveness refers to the property that the ASM is free of deadlocks while safety refers to the general absence of non-desired states in the ASM. A logical treatment of questions of liveness and safety can be seen as an indicator for the correctness of a designated ASM.

Consider for example the first ASM example of the previous section (as represented in figure \ref{fig:example1}). As mentioned before this ASM cannot be considered as free from deadlocks as a rejection of the request of \emph{compB} for the production plan of \emph{compA} does have not a solution in the ASM. The resulting deadlock however can be described by the following formula over the VDM++ specification which is underlying in the ASM:
$$\varphi := \exists \mathit{ord}. \exists \mathit{HA}. \exists \mathit{PP}. (\mathit{createHA}(\mathit{compB},\mathit{ord}) = \mathit{HA}$$
$$\wedge \mathit{createPP}(\mathit{ord}, \mathit{HA}) = \mathit{PP} \wedge \mathit{requestInfo}(\mathit{compB},\mathit{coal},\mathit{PP}) = \mathit{DENY}$$
Due to the general concept of AMSs with their freedom of choice in their underlying specifications and due to the expressiveness of the describle formulas there cannot exist a general algorithm for the test of satisfiability of formulas in given ASMs. However, this does not imply that such a test is impossible to define in any case. The structure of our example allows for a testing of the above formula $\varphi$ in the ASM as the number of states to consider is finite. It is obvious that the only situation in which $\varphi$ can be considered as a true formula is when calling the \emph{signoff()}-function after states $i_6$ and $i_10$. Thus a logical consideration could reveal the possible deadlock situation \emph{before} an explicit test run of the ASM has to take place. Doing that the ASM can be considered as not fulfilling the property of liveness by logical considerations.

As mentioned before the second ASM example (as represented in figure \ref{fig:example2}) of the previous section can be considered as free of deadlocks. Thus when attempting to test the validity of the above formula $\varphi$ in this second ASM the formula occurs to be false in all states and thus the ASM can be considered as satisfying the liveness property. Logical considerations of safety properties include all the testing for states of the ASM which are non-desired in the aspired specification of the ASM. Consider for example the following formula $\psi$:
$$\psi := \exists \mathit{ord}. \exists \mathit{HA}. \exists \mathit{PP}. (\mathit{createHA}(\mathit{compB},\mathit{ord}) = \mathit{HA}$$
$$\wedge \mathit{createPP}(\mathit{ord}, \mathit{HA}) = \mathit{PP} \wedge \mathit{requestInfo}(\mathit{compB},\mathit{coal},\mathit{PP}) = \mathit{PERMIT}$$
$$\to \mathit{sendErr}(\mathit{compB},\mathit{coal},\mathit{PP}))$$
This formula describes a situation in which the entire process is working as intended, but the system nevertheless sends an error to \emph{compA}. It is obvious that this formula $\psi$ describes an error state of the ASM which shall not occur as a possible state. Thus when testing the validity of $\psi$ in the ASM and find $\psi$ to be false in any state we can guarantee a priori that this non-desired state cannot occur in the ASM run.

Besides the mentioned aspects on liveness and safety there are further benefits of a logical consideration of ASMs (like for example consequence relations or other meta-concepts) which lead to more deep analysis of their structure and logical conditions.

\section{Conclusion and open questions}
In this paper we have presented a model in the formal specification language VDM, which may describe dynamic coalitions together with privacy and access control aspects. Our contribution here is the integration of two existing modeling approaches for dynamic coalitions and XACML access control policies respectively. Our next steps in this area will consider context-sensitive XACML as well as \emph{role based access control (RBAC)} for both of which OASIS-profiles exist. Both extensions to the basic XACML-concept are of high significance in most application fields, i.e. in the medical sector as the main field of interest of the authors.

Furthermore we have argued by our two examples of ASMs that the ASM method with its logical considerations can be of great benefit for a formal treatment of access-control-sensitive processes in a dynamic coalition. The VDM-structures serve as the basic states on which the ASMs operate. Here the examples are meant as a first indicator for the usefulness of a formal treatment as provided by the ASM method. However, it remains to define and study a general theory for a formal treatment of access control in dynamic coalitions using the ASM method which is general enough to cover a wide range of access control situations. These studies are part of ongoing and soon to be published work especially of the second author of the present paper.

\nocite{*}
\bibliographystyle{eptcs}
\bibliography{paper}

\begin{thebibliography}{10}
\providecommand{\bibitemdeclare}[2]{}
\providecommand{\urlprefix}{Available at }
\providecommand{\url}[1]{\texttt{#1}}
\providecommand{\href}[2]{\texttt{#2}}
\providecommand{\urlalt}[2]{\href{#1}{#2}}
\providecommand{\doi}[1]{doi:\urlalt{http://dx.doi.org/#1}{#1}}
\providecommand{\bibinfo}[2]{#2}

\bibitemdeclare{inproceedings}{BabSarrouh11}
\bibitem{BabSarrouh11}
\bibinfo{author}{Sebastian Bab} \& \bibinfo{author}{Nadim Sarrouh}
  (\bibinfo{year}{2011}): \emph{\bibinfo{title}{Formale Modellierung von
  Access-Control-Policies in Dynamischen Koalitionen}}.
\newblock In: {\sl \bibinfo{booktitle}{GI Proceedings, Informatik 2011 -
  Informatik schafft Communities}}, p. \bibinfo{pages}{402}.

\bibitemdeclare{inproceedings}{Bocchi09}
\bibitem{Bocchi09}
\bibinfo{author}{Laura Bocchi}, \bibinfo{author}{Jos{\'e}~Luiz Fiadeiro},
  \bibinfo{author}{Noor Rajper} \& \bibinfo{author}{Stephan Reiff-Marganiec}
  (\bibinfo{year}{2009}): \emph{\bibinfo{title}{Structure and Behaviour of
  Virtual Organisation Breeding Environments}}.
\newblock In: {\sl \bibinfo{booktitle}{FAVO}}, pp. \bibinfo{pages}{26--40}.
\newblock \urlprefix\url{http://dx.doi.org/10.4204/EPTCS.16.3}.

\bibitemdeclare{book}{Boe03}
\bibitem{Boe03}
\bibinfo{author}{Egon B\"orger} \& \bibinfo{author}{Robert St\"ark}
  (\bibinfo{year}{2003}): \emph{\bibinfo{title}{Abstract State Machines: A
  Method for High-Level System Design and Analysis}}.
\newblock \bibinfo{publisher}{Springer Verlag}.

\bibitemdeclare{techreport}{Bryans08}
\bibitem{Bryans08}
\bibinfo{author}{J.~W. Bryans}, \bibinfo{author}{J.~S. Fitzgerald},
  \bibinfo{author}{D.~Greathead}, \bibinfo{author}{C.~B. Jones} \&
  \bibinfo{author}{R.~J. Payne} (\bibinfo{year}{2008}): \emph{\bibinfo{title}{A
  Dynamic Coalitions Workbench: Final Report}}.
\newblock \bibinfo{type}{Technical Report}, \bibinfo{institution}{Newcastle
  University}.

\bibitemdeclare{techreport}{Bryans0601}
\bibitem{Bryans0601}
\bibinfo{author}{J.~W. Bryans}, \bibinfo{author}{J.~S. Fitzgerald},
  \bibinfo{author}{C.~B. Jones}, \bibinfo{author}{I.~Mozolevsky},
  \bibinfo{author}{Jeremy~W. Bryans}, \bibinfo{author}{John~S. Fitzgerald},
  \bibinfo{author}{Cliff~B. Jones} \& \bibinfo{author}{Igor Mozolevsky}
  (\bibinfo{year}{2006}): \emph{\bibinfo{title}{Dimensions of dynamic
  coalitions}}.
\newblock \bibinfo{type}{Technical Report}.

\bibitemdeclare{techreport}{Bryans0601a}
\bibitem{Bryans0601a}
\bibinfo{author}{J.~W. Bryans}, \bibinfo{author}{J.~S. Fitzgerald} \&
  \bibinfo{author}{P.~Periorellis} (\bibinfo{year}{2006}):
  \emph{\bibinfo{title}{Model Based Analysis and Validation of Access Control
  Polcies}}.
\newblock \bibinfo{type}{Technical Report}, \bibinfo{institution}{Newcastle
  University, School of Computing Science}.

\bibitemdeclare{incollection}{Bryans07}
\bibitem{Bryans07}
\bibinfo{author}{Jeremy Bryans} \& \bibinfo{author}{John Fitzgerald}
  (\bibinfo{year}{2007}): \emph{\bibinfo{title}{Formal Engineering of XACML
  Access Control Policies in VDM++}}.
\newblock In \bibinfo{editor}{Michael Butler}, \bibinfo{editor}{Michael
  Hinchey} \& \bibinfo{editor}{Mar{\'\i}a Larrondo-Petrie}, editors: {\sl
  \bibinfo{booktitle}{Formal Methods and Software Engineering}}, {\sl
  \bibinfo{series}{Lecture Notes in Computer Science}} \bibinfo{volume}{4789},
  \bibinfo{publisher}{Springer Berlin / Heidelberg}, pp.
  \bibinfo{pages}{37--56}.
\newblock \urlprefix\url{http://dx.doi.org/10.1007/978-3-540-76650-6_4}.

\bibitemdeclare{techreport}{Bryans06}
\bibitem{Bryans06}
\bibinfo{author}{Jeremy Bryans}, \bibinfo{author}{John~S. Fitzgerald},
  \bibinfo{author}{Cliff~B. Jones} \& \bibinfo{author}{Igor Mozolevsky}
  (\bibinfo{year}{2006}): \emph{\bibinfo{title}{Formal Modelling of Dynamic
  Coalitions, with an Application in Chemical Engineering}}.
\newblock \bibinfo{type}{Technical Report}.
\newblock \urlprefix\url{http://dx.doi.org/10.1109/ISoLA.2006.21}.

\bibitemdeclare{book}{Fitzgerald2005}
\bibitem{Fitzgerald2005}
\bibinfo{author}{John Fitzgerald}, \bibinfo{author}{Peter~Gorm Larsen},
  \bibinfo{author}{Paul Mukherjee}, \bibinfo{author}{Nico Plat} \&
  \bibinfo{author}{Marcel Verhoef} (\bibinfo{year}{2005}):
  \emph{\bibinfo{title}{Validated Designs For Object-oriented Systems}}.
\newblock \bibinfo{publisher}{Springer-Verlag TELOS}, \bibinfo{address}{Santa
  Clara, CA, USA}.

\bibitemdeclare{inproceedings}{Fitzgerald06}
\bibitem{Fitzgerald06}
\bibinfo{author}{John~S. Fitzgerald} \& \bibinfo{author}{Peter~Gorm Larsen}
  (\bibinfo{year}{2006}): \emph{\bibinfo{title}{Triumphs and Challenges for
  Model-Oriented Formal Methods: The VDM$^{\mbox{++}}$ Experience (Abstract)}}.
\newblock In: {\sl \bibinfo{booktitle}{ISoLA}}, pp. \bibinfo{pages}{1--4}.
\newblock \urlprefix\url{http://dx.doi.org/10.1109/ISoLA.2006.33}.

\bibitemdeclare{book}{FitzgeraldLarsen2009}
\bibitem{FitzgeraldLarsen2009}
\bibinfo{author}{John~S. Fitzgerald} \& \bibinfo{author}{Peter~Gorm Larsen}
  (\bibinfo{year}{2009}): \emph{\bibinfo{title}{Modelling Systems - Practical
  Tools and Techniques in Software Development (2. ed.)}}.
\newblock \bibinfo{publisher}{Cambridge University Press}.

\bibitemdeclare{proceedings}{Gure00}
\bibitem{Gure00}
\bibinfo{editor}{Yuri Gurevich}, \bibinfo{editor}{Philipp~W. Kutter},
  \bibinfo{editor}{Martin Odersky} \& \bibinfo{editor}{Lothar Thiele}, editors
  (\bibinfo{year}{2000}): \emph{\bibinfo{title}{Abstract State Machines, Theory
  and Applications, International Workshop, ASM 2000, Monte Verit{\`a},
  Switzerland, March 19-24, 2000, Proceedings}}. {\sl \bibinfo{series}{Lecture
  Notes in Computer Science}} \bibinfo{volume}{1912},
  \bibinfo{publisher}{Springer}.

\bibitemdeclare{inproceedings}{Haidar09}
\bibitem{Haidar09}
\bibinfo{author}{Ali~Nasrat Haidar}, \bibinfo{author}{P.~V. Coveney},
  \bibinfo{author}{Ali~E. Abdallah}, \bibinfo{author}{Peter Y.~A. Ryan},
  \bibinfo{author}{B.~Beckles}, \bibinfo{author}{J.~M. Brooke} \&
  \bibinfo{author}{M.~A.~S. Jones} (\bibinfo{year}{2009}):
  \emph{\bibinfo{title}{Formal Modelling of a Usable Identity Management
  Solution for Virtual Organisations}}.
\newblock In: {\sl \bibinfo{booktitle}{FAVO}}, pp. \bibinfo{pages}{41--50}.
\newblock \urlprefix\url{http://dx.doi.org/10.4204/EPTCS.16.4}.

\bibitemdeclare{inproceedings}{Klueber98}
\bibitem{Klueber98}
\bibinfo{author}{R.~Klueber} (\bibinfo{year}{1998}): \emph{\bibinfo{title}{A
  framework for virtual Organizing}}.
\newblock In: {\sl \bibinfo{booktitle}{VoNet Workshop}}.

\bibitemdeclare{article}{Koshutanski10}
\bibitem{Koshutanski10}
\bibinfo{author}{Hristo Koshutanski} \& \bibinfo{author}{Antonio Ma{\~n}a}
  (\bibinfo{year}{2010}): \emph{\bibinfo{title}{Interoperable semantic access
  control for highly dynamic coalitions}}.
\newblock {\sl \bibinfo{journal}{Security and Communication Networks}}
  \bibinfo{volume}{3}(\bibinfo{number}{6}), pp. \bibinfo{pages}{565--594}.
\newblock \urlprefix\url{http://dx.doi.org/10.1002/sec.148}.

\bibitemdeclare{article}{Larsen96}
\bibitem{Larsen96}
\bibinfo{author}{Peter~Gorm Larsen}, \bibinfo{author}{John Fitzgerald} \&
  \bibinfo{author}{Tom Brookes} (\bibinfo{year}{1996}):
  \emph{\bibinfo{title}{Applying Formal Specification in Industry}}.
\newblock {\sl \bibinfo{journal}{IEEE Softw.}} \bibinfo{volume}{13}, pp.
  \bibinfo{pages}{48--56}.
\newblock \urlprefix\url{http://doi.ieeecomputersociety.org/10.1109/52.493020}.

\bibitemdeclare{article}{Lethbridge01}
\bibitem{Lethbridge01}
\bibinfo{author}{Nick Lethbridge} (\bibinfo{year}{2001}):
  \emph{\bibinfo{title}{An I-Based Taxonomy of Virtual Organisations and the
  Implications for Effective Management}}.
\newblock {\sl \bibinfo{journal}{Informing Science}} \bibinfo{volume}{4 No 1}.

\bibitemdeclare{inproceedings}{McGinnis09}
\bibitem{McGinnis09}
\bibinfo{author}{Jarred McGinnis}, \bibinfo{author}{Kostas Stathis} \&
  \bibinfo{author}{Francesca Toni} (\bibinfo{year}{2009}):
  \emph{\bibinfo{title}{A Formal Framework of Virtual Organisations as Agent
  Societies}}.
\newblock In: {\sl \bibinfo{booktitle}{FAVO}}, pp. \bibinfo{pages}{1--14}.
\newblock \urlprefix\url{http://dx.doi.org/10.4204/EPTCS.16.1}.

\bibitemdeclare{book}{reisig10}
\bibitem{reisig10}
\bibinfo{author}{Wolfgang Reisig} (\bibinfo{year}{{2010}}):
  \emph{\bibinfo{title}{{Petrinetze: Modellierungstechnik, Analysemethoden,
  Fallstudien}}}.
\newblock \bibinfo{series}{{Leitf{\"a}den der Informatik}},
  \bibinfo{publisher}{{Vieweg+Teubner}}.
\newblock \bibinfo{note}{{248 pages; ISBN 978-3-8348-1290-2}}.

\end{thebibliography}
\end{document}